\documentclass[sigconf,natbib=false,timestamp=false]{acmart}

\usepackage[utf8]{inputenc}

\usepackage[abbreviate=true, dateabbrev=true, isbn=false, doi=false, urldate=comp, url=false, maxbibnames=9, backref=false, backend=biber, bibencoding=utf8, style=numeric-comp, language=american, giveninits=true, sortcites=true, sorting=none]{biblatex}
\usepackage{time}
\usepackage{listings}
\RequirePackage[l2tabu, orthodox]{nag}

\usepackage[T1]{fontenc}

\usepackage{csvsimple}

\usepackage{dsfont}
\usepackage{amsmath}
\usepackage{amsfonts}
\usepackage{amssymb}
\usepackage{amsbsy}
\usepackage{mathrsfs}

\PassOptionsToPackage{lowtilde}{url}

\usepackage{paralist}

\usepackage{url}

\usepackage{xcolor}

\usepackage{multirow}

\usepackage{booktabs}

\usepackage{csquotes}

\usepackage{tabularx}

\usepackage{graphicx}
\usepackage{color}

\usepackage[ruled,lined,linesnumbered]{algorithm2e}
\usepackage[noend]{algorithmic}

\usepackage{array}

\usepackage{hhline}

\usepackage{verbatim}

\usepackage[export]{adjustbox}
\usepackage{listings}

\usepackage{mathtools}

\usepackage{xspace}

\usepackage{todonotes}
\newif\ifdraft\drafttrue

\newcommand{\head}[1]{\par\noindent\textbf{#1:}\xspace}

\usepackage{booktabs}
\usepackage{csvsimple}
\usepackage{graphicx}

\newcommand{\CiscoNorway}{Cisco Systems Norway}

\begin{document}

\title{Improving Problem Identification via Automated Log Clustering using Dimensionality Reduction}

\author{Carl Martin Rosenberg}
\affiliation{%
  \institution{Simula Research Laboratory} 
  \city{Oslo, Norway}
}
\email{cmr@simula.no}

\author{Leon Moonen}
\affiliation{%
  \institution{Simula Research Laboratory}
  \city{Oslo, Norway}
}
\email{leon.moonen@computer.org}


\setcopyright{none}
\settopmatter{printacmref=false, printccs=false, printfolios=true}
\acmConference{Published in ESEM'18}{Proceedings of the 12th ACM/IEEE International Symposium on Empirical Software Engineering and Measurement}{October 2018, Article: 16, pp.\ 1–10,}
\acmPrice{}
\acmISBN{}
\acmDOI{}
\copyrightyear{\itshape\url{https://doi.org/10.1145/3239235.3239248}}

\begin{abstract}

\head{Background} 
Continuous engineering practices, such as continuous integration and continuous deployment, 
see increased adoption in modern software development. 
A frequently reported challenge for adopting these practices is the need to make
sense of the large amounts of data that they generate.

\head{Goal} 
We consider the problem of automatically grouping logs of runs that failed for the same underlying reasons,
so that they can be treated more effectively, and investigate the following questions:
(1) Does an approach developed to identify problems in \emph{system logs}
generalize to identifying problems in \emph{continuous deployment logs}?
(2) How does \emph{dimensionality reduction} affect the quality of automated log clustering?
(3) How does the criterion used for merging clusters in the clustering algorithm affect clustering quality?

\head{Method} 
We replicate and extend earlier work on clustering system log files to assess its generalization to continuous deployment logs.
We consider the optional inclusion of one of these dimensionality reduction techniques: 
Principal Component Analysis (PCA), Latent Semantic Indexing (LSI), and Non-negative Matrix Factorization (NMF). 
Moreover, we consider three alternative cluster merge criteria (Single Linkage, Average Linkage, and Weighted Linkage),
in addition to the Complete Linkage criterion used in earlier work.
We empirically evaluate the 16 resulting configurations on continuous deployment logs provided by our industrial collaborator.

\head{Results} 
Our study shows that 
(1) identifying problems in continuous deployment logs via clustering is feasible, 
(2) including NMF significantly improves \emph{overall} accuracy and robustness, 
and 
(3) Complete Linkage performs best of all merge criteria analyzed.

\head{Conclusions} 
We conclude that problem identification via automated log clustering is improved by including dimensionality reduction,
as it decreases the pipeline's sensitivity to parameter choice, 
thereby increasing its robustness for handling different inputs.

\end{abstract}


\maketitle
\renewcommand{\shorttitle}{Improving Problem Identif.\ via Automated Log Clustering w.\ Dimensionality Reduction}
\makeatletter
\fancyhead[LE]{\ACM@linecountL\@headfootfont\acmConference@shortname}%
\fancyhead[RO]{\@headfootfont\acmConference@shortname}%
\makeatother

\section{Introduction}
\label{sec:introduction}

\noindent
Continuous Engineering (CE) practices such as Continuous Integration (CI), 
Continuous Delivery (CDy), Continuous Deployment (CDt), and Continuous Release (CR), 
are increasingly adopted to meet the demand for incremental software development with rapid feedback.
Each of these practices can be characterized by their focus on creating short and automated cycles 
to give developers early feedback on potential issues and reduce risk by taking repeated incremental steps.
These cycles are triggered in response to new code being committed to the Version Control System, 
via periodic scripts, or simply when a developer demands it.

CI aims at automatically building and (unit) testing software changes multiple times a day, 
CDy extends it with automated acceptance testing and quality checking to ensure that a product is ready for deployment,
CDt adds automatic deployment to production-like hardware and deployment testing to CDy, and finally,
CR further adds automatically releasing the new software to the customers.\footnote{~%
There is a ongoing debate on the exact definitions and boundaries of these practices, 
but we will use the incremental CI $\subset$ CDy $\subset$ CDt $\subset$ CR definitions provided here.} 
When there is no need to distinguish, we will refer to them as CE.

In a traditional setup without CE, a developer or operator would manually build, test or deploy the system, 
observe the results and immediately react to them. 
With CE, on the other hand, it is possible to trigger a cycle that runs all necessary commands and tracks their outcomes in a logfile, 
switch context to a new task, and defer inspection of the resulting logfile to some later time.
When the cycles are time consuming, this increases developer productivity.
However, it also easily creates a situation where unprocessed results accumulate. 
This is not necessarily a problem, as long as the entire cycle is made to run correctly before the product is released.
However, the vast amounts of results that can accumulate may interfere with this last goal.

Indeed, a frequently reported challenge for adopting CE practices concerns 
the systematic and integrated analysis of the wealth of data resulting from the automated build, test, 
and deployment processes~\cite{Laukkanen2017,Hilton2017,Brandtner2015a,Brandtner2015b,Debbiche2014,Olsson2012}.  
A recent literature review identifies the lack of transparency and awareness regarding test and build results as one of the main threats to adopting CDy, 
together with the need for measures that improve coordination and collaboration on addressing such results~\cite{Shahin2017}.
As similar challenges were brought up by our industrial collaborator, we set out to investigate techniques to automatically support such coordination.

\head{Goal}
We consider the challenge of automatically grouping logs of CE runs that failed for the same underlying reasons.
The idea is that instead of having to investigate all individual results, 
automated log clustering would allow for a more systematic, coordinated approach where one investigates a representative for a group of failed runs, 
addresses the issues that caused failure, and can reasonably expect that the issues of logs that failed for corresponding reasons have been accounted for 
(which will be checked by following runs).
To this end, we replicate and extend earlier work by Lin et al.~\cite{Lin2016} that proposes an automatic clustering pipeline for system log sequences, 
so that an system operator diagnosing an issue only needs to investigate a few representative sequences.

\head{Contributions}
The main contributions of this work can be summarized as follows:
(1) We conduct a replication study to investigate to what extent an approach developed to identify problems in system logs
generalizes to the context of identifying problems in CDt logs.
(2) We extend the original study in two ways:
(2a) by investigating how the optional inclusion of one of three dimensionality reduction techniques affects the quality of automated log clustering, and
(2b) by investigating three additional alternative criteria for merging clusters in the Hierarchical Agglomerative Clustering algorithm that it uses.
(3) We empirically investigate the impact of the 16 resulting configurations on the quality of clustering CDt logs provided by our industrial collaborator. 
(4) We analyze and discuss how dimensionality reduction helps to lower the pipeline's sensitivity to parameter choices, 
thereby increasing its robustness for handling different inputs.

\head{Overview}
The remainder of this paper is organized as follows:
Section~\ref{sec:background} discusses Lin's approach, and
Section~\ref{sec:variations} presents the variations that we investigate.
Section~\ref{sec:experimental_design} describes the setup of our empirical investigation, 
whose results are presented and discussed in Section~\ref{sec:results}.
Section~\ref{sec:related_work} presents related work, and
Section~\ref{sec:conclusion} provides some concluding remarks.

\medskip

\section{Background}
\label{sec:background}

\noindent
The baseline for our work is LogCluster by Lin et al.~\cite{Lin2016}, 
a technique developed for analyzing system logs from online services. 
This section presents the main parts of the LogCluster pipeline, 
and the next section describes our adaptions to it.

Conceptually, LogCluster groups logs into clusters, selects a representative log for each cluster, and presents these to the developer. 
To perform the clustering, the logs must first be preprocessed and represented in a manner that is amenable to clustering. 
In LogCluster, this comprises a \emph{log abstraction} step that represents each log as a sequence of \emph{events}, 
and a \emph{log vectorization} step that uses \emph{inverse document frequency} and \emph{event contrast} to give higher priority to rare or specific events. 
We now present these steps in more detail.\footnote{~Note that we do not adopt the LogCluster scheme~\cite{Lin2016} wholesale. 
In particular, the use of a database of previously seen Log Sequences to save computational costs and increase performance 
is an orthogonal extension that is beyond the scope of this paper.}

\head{Log abstraction}
This step removes runtime-specific information that is assumed to create artificial differences between logs. 
For example, most logs report timestamps that add such differences.
Abstraction represents all timestamps using the same token, 
so the algorithm can focus on more salient aspects of the logs instead. 
Lin et al.~\cite{Lin2016} use the automatic abstraction mechanism developed by Fu et al.~\cite{Fu2009a}. 
In our work, we use a log abstraction tool provided by our industrial collaborator, \CiscoNorway{}.

Next, each log is represented as a sequence of \emph{event identifiers}. 
This assumes that the log file is a series of event reports delineated in a predictable manner, 
for example a newline followed by a date, time and timezone designator. 
After abstraction  of runtime-specific information, many of these event descriptions will be identical.
Thus, by giving each of these identical event descriptions the same unique identifier, 
logs can be compactly represented as sequence of event identifiers, 
which benefits the next steps of the pipeline.

\head{Log Vectorization}
The logs must be represented in a way that enables reasoning about the differences between them. 
A common representation strategy for text documents is the \emph{bag-of-words} model~\cite{manning2008introduction}.
In this model, every document is represented as a vector, 
the length of which is determined by the number of unique words in the corpus to be represented, 
as each cell in the vector represents a specific word.
For a document, each cell in the vector tracks the number of times its corresponding word occurs in that document. 
LogCluster employs a similar strategy, but uses the event identifiers that were uncovered in the log abstraction step instead of words. 

\head{Event frequency weighting}
When employing a bag-of-words-style representation strategy, 
it is common to transform the obtained vectors with an \emph{inverse document frequency scheme}~\cite{manning2008introduction}.
The goal is to give higher weight to words (or in LogCluster's case, events) that are rare in the documents to be clustered, 
and proportionally lower weight to frequently occurring words.
In LogCluster~\cite{Lin2016}, the event frequency weight $w_{f}(e)$ of an event $e$ is given by:
$w_{f}(e) = S(\log{\frac{N}{n_e}}) $
where $N$ is the number of logs to be clustered, $n_e$ is the number
of logs where the event $e$ appears, and $S$ is the Sigmoid function
$1/(1+\exp[{-x}])$
which normalizes the vector by ensuring that all values are between 0 and 1.

\head{Contrast-based event weighting}
Given separate sets of "interesting" logs (in our case, logs
from failing CE runs) and "uninteresting" logs (i.e. the passing CE runs),
Lin et al.~\cite{Lin2016} propose to give a lower weight to those
events which occur in both the interesting and uninteresting sets. They
refer to this scheme as \emph{contrast-based event weighing}, and it
works as follows: First, determine the set of events $\Delta S$ that
only occur in the interesting set of logs, and then use $\Delta S$ to
assign a \emph{contrast weight} $w_{con}(e)$ to each event such that
$w_{con}(e)=1$ if $e \in \Delta S$, and $w_{con}(e)=0$ otherwise.

\noindent
The final weight of each event is determined by combining
contrast weight with the inverse document frequency weight.
LogCluster assigns equal weight
to the inverse document frequency weight and the contrast-based
weight, so that the final event weight $w(e)$ becomes
$ w(e) = 0.5 * w_{con}(e) + 0.5 * w_{f}(e) $.

\head{Clustering and selecting representatives}
LogCluster employs the \emph{cosine distance} dissimilarity metric and 
Hierarchical Agglomerative Clustering (HAC) to cluster the logs. 
The clustering stops when the calculated merge distance between two candidate cluster exceeds a given threshold $\theta$, which LogCluster sets to $0.5$.
Finally, a representative for each cluster is chosen by finding the log with the smallest average cosine distance to the other logs in the cluster.

\section{Variations on LogCluster}
\label{sec:variations}

\begin{figure}[t]\centering
\vspace{.5ex}
\includegraphics[width=.99\columnwidth]{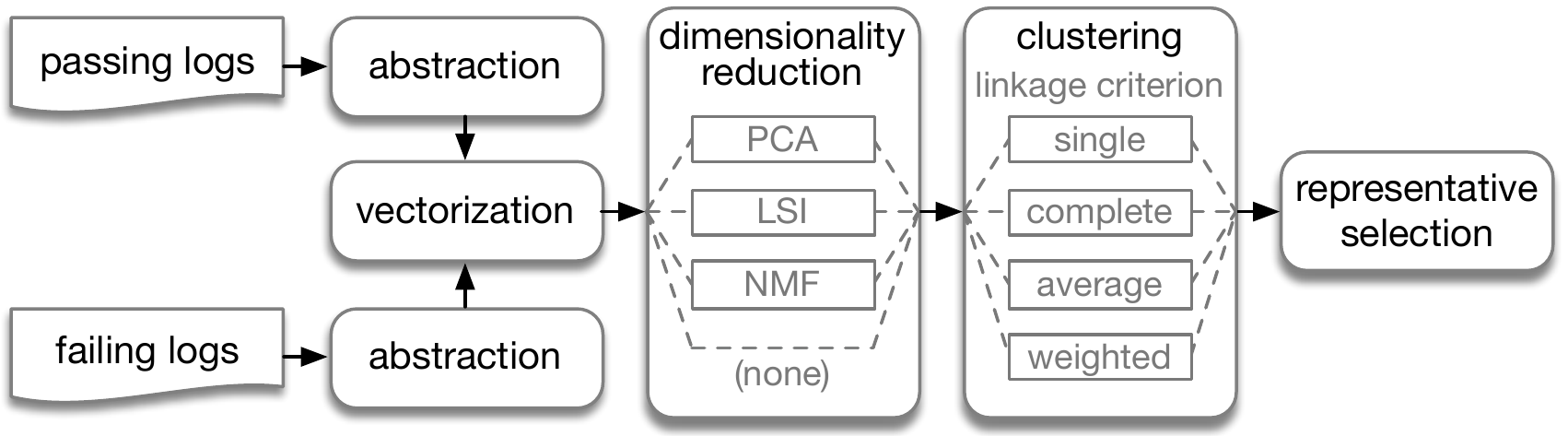}
\vspace*{-1ex}
\caption{Overview of the variations in the clustering pipeline investigated in this paper.}
\label{fig:pipeline}
\vspace*{-4ex}
\end{figure}

This section motivates and discusses the variations on the LogCluster pipeline (Section~\ref{sec:background}) that we investigate in our work. A high level overview of the variations is presented in Section~\ref{fig:pipeline}.

\subsection{Dimensionality reduction}

Recall that if there are $n$ unique events in a collection of logs,
each log will be represented by an $n$-dimensional vector. 
For diverse collections, $n$ can quickly grow enormous. 
Since the clustering algorithm reasons about distances between vectors in an $n$-dimensional space, 
the algorithm is vulnerable to Bellman's \emph{curse of dimensionality}: 
As the number of dimensions $n$ grows, 
it becomes increasingly difficult to properly discriminate between points~\cite{Aggarwal2015,Hinneburg2000}.

A strategy to combat the curse of dimensionality is to apply \emph{dimensionality reduction}, 
a transformation of the data that exploits correlations among dimensions 
to create a reduced space that capture most of the variance in the data~\cite{Aggarwal2015}.
Note that dimensionality reduction can improve the ability to discriminate between points, 
but there is also a risk of removing too much information.

We investigate the impact of applying one of the following dimensionality reduction techniques: 
Principal Component Analysis (PCA), Latent Semantic Indexing (LSI) or Non-Negative Matrix
Factorization (NMF). 
While PCA is heavily used as a dimensionality reduction technique in all domains, 
LSI and NMF have shown particular promise in text applications.
LSI has been shown to be 
particularly effective on text collections plagued by \emph{polysemy} (when words have multiple meanings) 
and \emph{synonymy} (when several words have the same meaning)~\cite{manning2008introduction}. 
NMF has been shown to be especially effective when the data is \emph{sparse} and \emph{non-negative}, 
which is the case for textual data in the bag-of-words model~\cite{Aggarwal2015}.
All entries in a bag-of-words vector will be either zero or some positive value as they track word frequencies.
Since a document typically only contains a fraction of the words occurring throughout the collection of documents, 
the vectors tend to be sparse, i.e. most cells will be 0.

\head{Estimating the number of components}\label{compestim}
To use these dimensionality reduction techniques, 
one must typically specify a number of components $k$ such that the data gets reduced to a $k$-dimensional space, 
and one of the outcomes is an estimate of the \emph{explained variance} with respect to the original data.
For PCA, each component has a known contribution to the overall explained variance (evar),
and components are typically ordered by decreasing contribution.
This allows us to determine how many components are needed to reach a desired explained variance, 
by first performing an unconstrained PCA, 
and then iteratively accumulating components (and evar contributions) until the desired value is exceeded.

We conjecture that this PCA-based computation also provides an appropriate estimate for the components 
needed by LSI and NMF to achieve the same explained variance. 
We aim at an explained variance of 80\%, a value commonly used in the literature~\cite{Korenius2007},
use the PCA and LSI implementations from scikit-learn~\cite{scikit-learn}, 
and the NMF implementation provided by Nimfa~\cite{Zitnik2012}.

\newcommand{\criterion}[1]{\par\noindent\textbf{#1}}

\subsection{Alternative merge criteria in HAC}
\label{complete}

Another point of variation that we investigate is the criterion that is used to merge clusters in the
clustering step of LogCluster.
The Complete Linkage merge criterion that is used in LogCluster is known to be sensitive to outliers, 
which could prevent creation of the most intuitive clusters~\cite[Sec.\ 17.2]{manning2008introduction}. 
For this reason, we investigate the impact of using one of the following alternative merge criteria:
Single Linkage, Average Linkage, and Weighted Linkage. 
We give a short overview of the intuitions behind the various merge criteria,
and refer to M\"{u}llner~\cite{Mullner2011} for details: 
\criterion{Single Linkage} 
merges the two distinct clusters for which the \emph{smallest} distance between a pair of elements from either cluster is the \emph{global minimum} for all pairs of clusters. 
\criterion{Complete linkage}
merges the two distinct clusters for which the \emph{maximum distance} between a pair of elements from either cluster is the \emph{global minimum} for all pairs of clusters. 
\criterion{Average Linkage (UPGMA)}
merges the two distinct clusters for which the \emph{average distance} between \emph{all pairs of elements} from either cluster is the \emph{global minimum} for all pairs of clusters. 
\criterion{Weighted Linkage (WPGMA)}
merges a previously merged cluster $A \cup B$ and merge candidate $C$ for which the \emph{average distance} between sub-cluster pairs $(A,C)$ and $(B,C)$ is the \emph{global minimum}.

Our experiments use the HAC implementations from SciPy~\cite{Jonesa}, which implement M\"{u}llner's algorithm and merge criteria~\cite{Mullner2011}. 

\section{Experimental Design}
\label{sec:experimental_design}

We conduct an empirical study to assess whether logs of CE runs that failed for the same underlying reasons can be 
automatically grouping using automated log clustering, 
and what improvements  can be achieved in terms of accuracy and robustness.
Specifically, we set out to answer the following research questions:
\begin{compactenum}[\bfseries RQ1]
\item Can the application of LogCluster~\cite{Lin2016} be generalized to identifying problems in \emph{continuous deployment log files}?
\smallskip
\item To what extent does applying dimensionality reduction impact the results of  automated log clustering?
\smallskip
\item To what extent does the merge criterion in HAC impact  the results of  automated log clustering?
\end{compactenum}
Next, we detail the experimental design to answer these questions.

\subsection{Datasets}
\label{sec:dataset}

We use a dataset provided by \CiscoNorway{} consisting of CDt logs and associated meta-data for 18 different comprehensive integration tests.
The logs are loosely structured execution outputs that capture the process of building, deploying and testing various scenarios, 
not unlike the output of running a make command.
Table~\ref{tbl:dataset-summaries} summarizes the main features of the data.

\begin{table}[t]\centering
\caption{Overview of datasets used in this investigation.}
\label{tbl:dataset-summaries}\small
\vspace*{-2.5ex}
\begin{filecontents*}{dataset_metadata.csv}
dataset,failing,passing,dimensions,avg. dims. a/red.
1,45,4189,164,9.50
2,102,4182,192,8.00
3,103,1315,850,15.00
4,8,4277,70,1.00
5,41,297,118,5.50
6,23,2421,352,6.50
7,50,1591,71,2.50
8,78,3213,605,2.00
9,52,461,177,8.00
10,342,3152,201,17.83
11,15,3816,23,2.00
12,74,2811,133,9.00
13,106,572,181,8.00
14,60,613,162,9.50
15,140,3247,204,15.50
16,132,3238,213,16.50
17,36,530,201,7.50
18,29,344,171,6.00
\end{filecontents*}
\csvautobooktabular{dataset_metadata.csv}\\
~\parbox{0.86\columnwidth}{The number of dimensions indicates the number of unique events in the dataset. 
The rightmost column shows the average number of dimensions after dimension reduction with PCA.}
\vspace*{-2.5ex}
\end{table}

The dataset is constructed so that every failing run is associated with one (and only one) known issue. 
Thus, we have a \emph{ground truth} for how the logs should be clustered together
so that each cluster only contains logs concerning the same issue. 
This allows us to evaluate the accuracy of proposed clusterings. 
The ground truth is derived from handcrafted regular expressions
developed by \CiscoNorway{} to identify whether a log concerns a known issue.

\subsection{Accounting for Parameter Sensitivity}
\label{sec:parameters}

Recall from Section~\ref{sec:background} how the contrast-based event weighting in LogCluster~\cite{Lin2016} 
assigns equal weight to the inverse document frequency weight and the contrast-based weight.  
We submit that it is unlikely that an equal split between the contrast
weight and the inverse document frequency weight is suitable for all
scenarios and inputs.
To include a wider range of splits in our study, 
we introduce the contrast parameter $\gamma$ such that the event weight $w(e)$ is given by:
$w(e) = \gamma * w_{con}(e) + (1-\gamma) * w_{f}(e)$.
Observe that the original LogCluster approach is equivalent to $\gamma = 0.5$. 

A second parameter to consider is the threshold $\theta$ used by the clustering algorithm to decide when to stop merging clusters. 
LogCluster sets this to $0.5$.
Also for this parameter we argue that it is unlikely that a single choice is suitable for all scenarios and inputs.

Low sensitivity to parameter choices implies the pipeline has increased robustness for handling different inputs, 
a desirable characteristic for an unsupervised approach such as automated log clustering~\cite{Aggarwal2015}.
To assess the sensitivity of different choices of dimensionality reduction and merge criteria
to changes in parameters $\theta$ and $\gamma$,
we execute each of the 16 configurations discussed in Section~\ref{sec:variations} on each of the 18 datasets with 
21 choices of $\gamma$ from $0$ up to and including $1$ in increments of $0.05$, and 
17 different choices of $\theta$ from $0.1$ up to and including $0.9$ in increments of $0.05$.
Thus, we execute each configuration (a choice of dimensionality reduction and merge criterion) a total of 357 times, 
each run representing a different choice of the $\gamma$ and $\theta$ parameters on a specific dataset.

\newcommand{\ours}{\triangle}
\newcommand{\ourm}{\triangle}
\newcommand{\lins}{\blacksquare}
\newcommand{\linm}{\blacksquare}
\newcommand{\ourrow}{$\ours$ & $\ourm$ & $\ours$ & $\ourm$ & $\ours$ & $\ourm$ & $\ours$ & $\ourm$ & $\ours$ & $\ourm$ & $\ours$ & }
\newcommand{\linrow}{$\ours$ & $\ourm$ & $\ours$ & $\ourm$ & $\lins$ & $\linm$ & $\lins$ & $\ourm$ & $\ours$ & $\ourm$ & $\ours$ & }

\begin{table*}[t]
\centering
\caption{Overview of our experimental design. Each configuration is used as \emph{treatment} (a row in the table), and every distinct choice of $\gamma$, $\theta$ and data-set as a \emph{block} (a column in the table). The earlier work by Lin et al.~\cite{Lin2016} is indicated as $\lins$, new data as $\ours$.}
\label{tbl:friedman-illustration}
\vspace*{-1.5ex}\small
\begin{tabular}{ccc|clllllllllll|l} %
\toprule
\multicolumn{2}{c}{configurations} &&&  \multicolumn{12}{c}{results (AMI)}                                                        \\ \midrule
dimension. & HAC       &&&  1 & \ldots & 18 & \ldots & 1 & \ldots & 18 & \ldots & 1 & \ldots & 18 & dataset \\
reduction    & merge       &&& 0.00 & \ldots & 0.05 & \ldots & 0.5 & \ldots & 0.5 & \ldots & 0.95 & \ldots & 1.0 & $\gamma$ \\
technique    & criterion   &&& 0.10 & \ldots & 0.15 & \ldots & 0.5 & \ldots & 0.5 & \ldots & 0.95 & \ldots & 1.0 & $\theta$ \\\midrule
\emph{none}  & complete    &&&  \linrow{} \\
             & single      &&&  \ourrow{} \\
             & average     &&&  \ourrow{} \\
             & weighted    &&&  \ourrow{} \\
PCA          & \ldots      &&&  \ourrow{} \\
LSI          & \ldots      &&&  \ourrow{} \\
NMF          & \ldots      &&&  \ourrow{} \\\bottomrule
\end{tabular}
\vspace*{-1ex}
\end{table*}

\subsection{Quality Measures}
\label{sec:qm}

We evaluate the performance of all runs with the various configurations using the following quality measures:

\head{Adjusted Mutual Information}
Lin et al.~\cite{Lin2016} use \emph{Normalized Mutual Information} (NMI)~\cite{manning2008introduction} to benchmark LogCluster. 
However, NMI has a systematic bias in favor of clustering algorithms that group data into many small clusters, 
as those are more likely to have many agreements solely due to chance~\cite{Vinh2009}.
For this reason, we use \emph{Adjusted Mutual Information} (AMI), which corrects for this bias~\cite{Vinh2009}. 
It is used to compare two ways of partitioning the input, in our case the ground truth and the clustering result. 
An AMI score can maximally be $1$, indicating a perfect correspondence between the proposed clustering and the ground truth. 
Conversely, an AMI near $0$ indicates that the proposed clustering performs as one would expect from a solution based on random
guessing.\footnote{~A caveat for inspecting our results is that an AMI score of $0.5$ does not correspond to random guessing, in contrast to many other accuracy measures (e.g., the F1 score).}

\head{Effort Reduction} 
We measure how much
less effort the operator has to exert after clustering by computing
an \emph{effort reduction} measure
$ER=1 - (\text{\emph{distinct proposed clusters}}/n)$
for each dataset, where $n$ is the number of logs to be
clustered. We also record the effort reduction a perfect solution would achieve
by computing $IER=1 - (\text{\emph{distinct ground-truth clusters}}/n)$.

\head{Homogeneity and Completeness} 
These represent two competing quality concerns~\cite{rosenberg2007v}. 
Homogeneity measures the extent to which members of a proposed cluster come from the same ground-truth class. 
Completeness, on the other hand, 
measures to what extent all members of a given ground-truth class are put in the same cluster. 
Perfect Homogeneity can trivially be achieved by putting each data-point in an individual cluster, 
but this solution will score low on Completeness. 
Conversely, perfect Completeness can be achieved by putting every data-point in a single cluster, 
but such a solution would obtain a very low Homogeneity score.
Thus, in order to achieve high accuracy a clustering algorithm must score
high on both Homogeneity and Completeness.

These four quality metrics gives us a nuanced view of a configuration's performance. 
If accuracy is low, we can determine whether it lacks Homogeneity,
Completeness or both, and we can assess whether it has a bias towards either Homogeneity and Completeness.
We expect Completeness to be heavily correlated with effort reduction, as both favor large clusters. 
Homogeneity, furthermore, serves as a risk indicator: 
A high Homogeneity score indicates that each log in a cluster can serve as a good representative for the issues in its cluster 
(i.e., few false positives or mis-clustered logs).

\subsection{Statistical Procedures} 

To compare the configurations, we first run a Friedman test~\cite{friedman1937}, 
as recommended by Dem\u{s}ar~\cite{Demsar2006}. 
We let each combination of dimensionality reduction technique and merge criterion represent a distinct \emph{treatment}, 
and every distinct combination of dataset, $\gamma$ and $\theta$ represent a \emph{block}, 
as illustrated in Table~\ref{tbl:friedman-illustration}. 

The Friedman test checks the null hypothesis that the treatments are equally effective 
against the alternative that at least one pair of treatments differ. 
With twenty-one choices of contrast $\gamma$, seventeen choices of threshold value $\theta$, 
18 different datasets and 16 competing configurations, 
our scheme has 16 different treatments and $18 \times 21 \times 17 = 6426$ blocks.
If the Friedman test rejects the null hypothesis of equal effectiveness at significance level $\alpha=0.05$, 
we proceed with a post-hoc analysis consisting of a paired Wilcoxon signed-rank test~\cite{wilcoxon1945} on each pair of treatments, 
as recommended by Benavoli et al.~\cite{Benavoli2016}. 
Thus, when comparing treatment $a$ against treatment $b$, 
the paired Wilcoxon test will for every block (i.e. choice of data-set, $\gamma$ and $\theta$) 
pair the measurement made for $a$ and the measurement made for $b$ on that block. 
We apply the the Pratt correction~\cite{Pratt1959} to the Wilcoxon signed-rank tests to handle ties, 
and control the family-wise error rate resulting from multiple comparisons with Holm's procedure~\cite{Holm1979}. 
We measure the effect size of each pairwise comparison in terms of Vargha-Delaney $A_{12}$ and $A_{21}$~\cite{Vargha2000}.

We use the \texttt{\small stats.wilcoxon} and \texttt{\small stats.friedmanchisquare} procedures from SciPy~\cite{Jonesa} 
to implement our statistical tests.

\section{Results and Discussion}
\label{sec:results}

This section first addresses RQ1 by assessing the performance of LogCluster on our dataset with respect to the quality measures discussed earlier
 (Section~\ref{sec:qm}). 
Next, we present the results of the simultaneous empirical evaluation of all 16 configurations that arise from the variations for both dimensionality reduction and merge criterion. 
Finally, we use these results to answer RQ2 and RQ3.

\begin{table}[b]\centering
\vspace*{-3ex}
\caption{Results for LogCluster's exact configuration}
\label{tbl:rq1results}
\vspace*{-2ex}\small
\begin{tabular}{@{}llllll@{}}
\toprule
Dataset         & AMI   & H     & C     & ER    & IER   \\ \midrule
1               & 0.383 & 0.775 & 0.537 & 0.733 & 0.844 \\
2               & 0.673 & 0.853 & 0.747 & 0.853 & 0.873 \\
3               & 0.201 & 0.676 & 0.251 & 0.883 & 0.951 \\
4               & 1.000 & 1.000 & 1.000 & 0.875 & 0.875 \\
5               & 0.085 & 0.171 & 0.135 & 0.927 & 0.927 \\
6               & 0.211 & 0.327 & 1.000 & 0.870 & 0.609 \\
7               & 0.676 & 0.705 & 1.000 & 0.920 & 0.900 \\
8               & 0.341 & 0.722 & 0.494 & 0.808 & 0.859 \\
9               & 0.623 & 0.708 & 0.815 & 0.846 & 0.827 \\
10              & 0.569 & 0.645 & 0.677 & 0.936 & 0.912 \\
11              & 1.000 & 1.000 & 1.000 & 0.933 & 0.933 \\
12              & 0.453 & 0.734 & 0.574 & 0.797 & 0.892 \\
13              & 0.200 & 0.539 & 0.317 & 0.887 & 0.887 \\
14              & 0.263 & 0.562 & 0.445 & 0.817 & 0.817 \\
15              & 0.734 & 0.855 & 0.808 & 0.836 & 0.857 \\
16              & 0.565 & 0.734 & 0.722 & 0.803 & 0.826 \\
17              & 0.628 & 0.780 & 0.796 & 0.722 & 0.722 \\
18              & 0.281 & 0.607 & 0.506 & 0.759 & 0.724 \\ \midrule
\textbf{Median} & 0.509 & 0.715 & 0.700 & 0.850 & 0.866 \\
\textbf{Mean}   & 0.494 & 0.689 & 0.657 & 0.845 & 0.846 \\ \bottomrule
\end{tabular}
\end{table}

\subsection{RQ1: Generalization of LogCluster to CE}

The performance of LogCluster is reported in Table~\ref{tbl:rq1results}.
We observe a median AMI of $0.509$ and a wide spread:
LogCluster achieves a perfect AMI score on datasets 4 and 11, 
but scores below $0.22$ on datasets 3, 5, 6 and 13. 
The perfect scores are obtained on small datasets (see Table~\ref{tbl:dataset-summaries}). 
On datasets 3, 8 and 13, LogCluster achieves relatively high Homogeneity, 
but very low Completeness. 
Conversely, dataset 6 yields perfect Completeness but rather low Homogeneity (0.327). 
Performance on dataset 5 is poor on all accounts.

For over two-thirds of the data, 
the obtained effort reduction (ER) is slightly below the ideal effort reduction (IER).
Exceptions are datasets 6, 7, 9, 10 and 18 where ER exceeds IER,
suggesting fewer clusters were created than required by the ground truth.

\head{Trade-offs}%
Recall that Homogeneity can be seen as a risk measure, 
reflecting the probability that a cluster element accurately represents all the issues of its cluster. 
Likewise, Completeness can serve as proxy for effort reduction, 
as Completeness rewards creating large clusters. 
Depending on preferences, a user could choose to adopt LogCluster in scenarios where the overall AMI score is low, 
but either the Homogeneity or Completeness score is high. 
For dataset 3 for example, a risk-averse user could elect to use LogCluster at a Homogeneity score of $0.676$, 
and accept that the effort reduction achieved is $88\%$ rather than $95\%$.
In this case one basically trades off dealing with a few more clusters for the assurance that they do not contain false positives.

\head{Answer to RQ1} Based on our study using data from our industrial partner, our answer to RQ1 is 
that the LogCluster approach generalizes to problem identification in continuous deployment logs.
Moreover, a user can make a trade-off between risk and effort reduction, for example by choosing to deal with a few more clusters 
for the assurance that they ate less likely to contain false positives.

\begin{figure}[b]\centering
\vspace*{-3ex}
\includegraphics[width=.95\columnwidth]{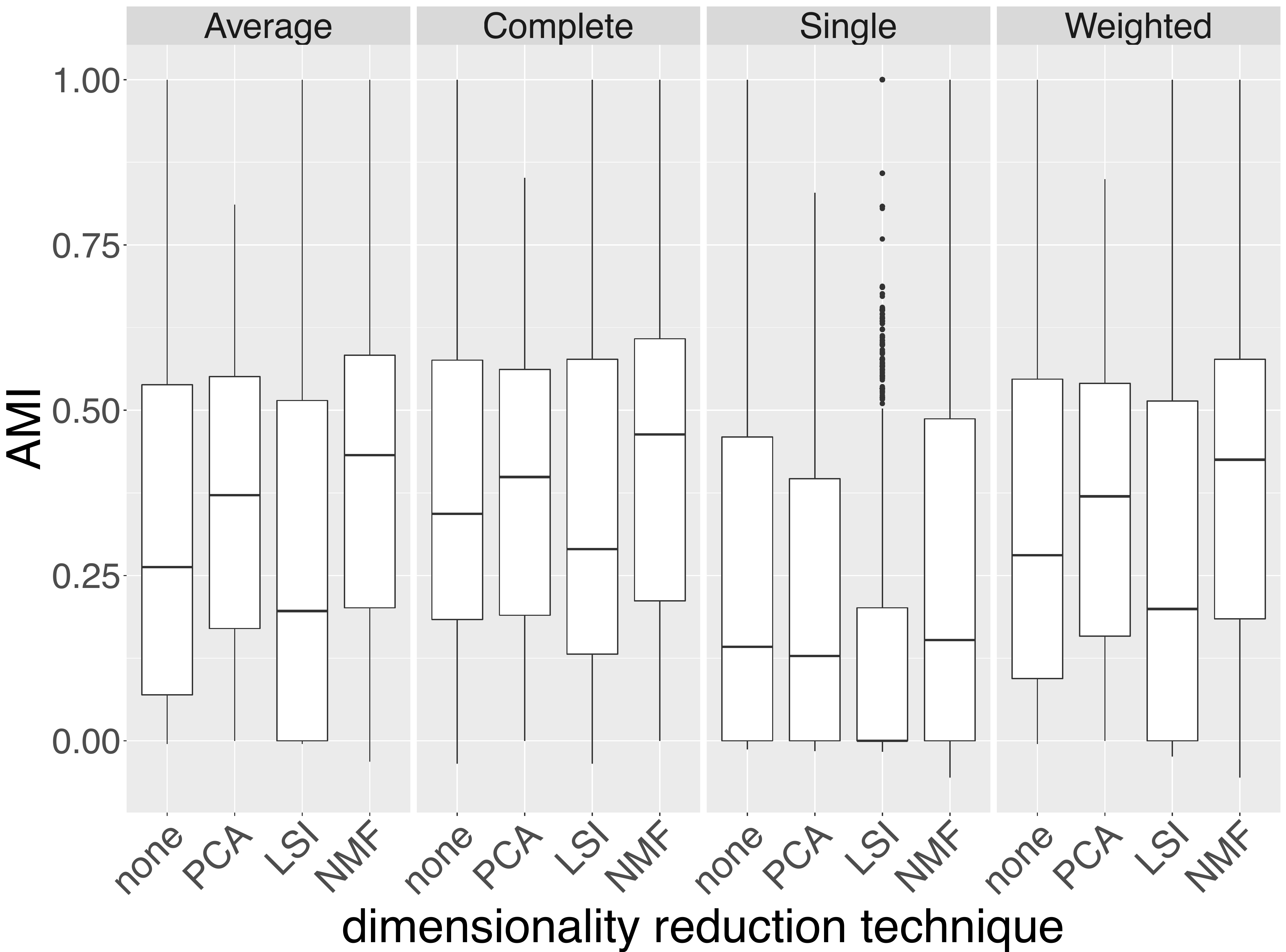}
\vspace*{-2ex}
\caption{Boxplots of AMI scores for each configuration.}
\label{fig:amiboxplots}
\end{figure}

\subsection{Analysis of LogCluster and its Variations}

Our blocked design yields 6426 results for each of the 16
configurations. Figure~\ref{fig:amiboxplots} shows boxplots of the AMI for
each configuration. The Friedman test reported a chi square statistic
of 19593.623 at a p-value lower than $2.2e-16$. We thus reject the
null hypothesis that the configurations are equally performant at
significance level $\alpha=0.05$, and proceed with the post-hoc
analysis.

\begin{table}[t]
\caption{Summary of post-hoc comparisons (A v. B) for which the null hypothesis could not be rejected.}\centering
\label{tbl:insignificantcomparisons}
\vspace*{-1.6ex}\small\setlength{\tabcolsep}{2.5pt}
\begin{tabular}{cc cc cccc}\toprule
\multicolumn{2}{c}{A}  & \multicolumn{2}{c}{B} \\
dimr & crit    & dimr & crit    & T        & VDA$_{AB}$  & VDA$_{BA}$  & p \\\midrule
NMF  & Average  & PCA  & Complete & 10026913 & 0.5411 & 0.4589 & 0.0450 \\
NMF  & Weighted & PCA  & Weighted & 10295318 & 0.5570 & 0.4430 & 0.8419 \\\bottomrule
\end{tabular}
\vspace*{-4ex}
\end{table}

Our post-hoc analysis consists of $\binom{16}{2}=120$ paired Wilcoxon Signed-Rank tests: 
One test for each combination of configurations. 
All comparisons were found to be significant after the Holm correction, 
except for the two comparisons shown in Table~\ref{tbl:insignificantcomparisons}. 
Due to the presence of insignificant comparisons we cannot impose a total order on the competing configurations. 
Table~\ref{tbl:condensedsummary} reports the variants by their Rank-Sum groups, 
including multiple alternatives in the same group when the Wilcoxon Signed-Rank test found no significant differences. 
We make the following observations from Table~\ref{tbl:condensedsummary}, 
which are corroborated by the boxplots in Figure~\ref{fig:amiboxplots}:
\begin{compactenum}[(1)]
\item NMF with Complete Linkage scores significantly better on both mean and median AMI 
      than all other considered configurations.

\item LSI is the worst performing dimensionality reduction technique. While
LSI with complete linkage scores better than \emph{no} dimensionality reduction with weighted
or average linkage, this is likely due to complete linkage rather than
LSI, as \emph{no} dimensionality reduction with complete linkage performs better. 

\item For any choice of dimensionality reduction technique, 
Complete linkage is the best performing cluster merge criterion.

\item Single linkage is the worst performing cluster merge criterion, 
regardless of dimensionality reduction technique.

\item Average and Weighted linkage have very similar performance characteristics, with
the strongest difference obtained when used in combination with NMF.

\item For all merge criteria besides single linkage, PCA and NMF
obtain a better result than \emph{no} dimensionality reduction and complete linkage, which
is LogCluster's configuration~\cite{Lin2016}.
\end{compactenum}

\head{Effect sizes}
For each pairwise comparison, we measure the effect size in terms of $VDA = max(A_{12}, A_{21})$. 
Recall that 118 of the 120 comparisons obtained statistically significant results, 
with the two remaining comparisons shown in Table~\ref{tbl:insignificantcomparisons}.
The majority of the 118 comparisons indicate small or small-to-medium effects.
Concretely, 84 of the 118 significant comparisons yielded small effect sizes at or below $0.64$,
while the remaining 34 comparisons exceeded this threshold.
Unsurprisingly, the large effects sizes occur when one of the six poorest performers 
(as indicated by Table~\ref{tbl:condensedsummary}) 
is compared to one of the ten best performers. 
When comparing the best performer (NMF + Complete) against the poorest performer (LSI + Single), 
we measure a VDA score of $0.796$ (a large effect).
We present the effect sizes of the more performant configurations with respect to the LogCluster baseline (no dimensionality reduction with complete linkage) in Table~\ref{tbl:bestvbaseline}. 
The improvements over the baseline are statistically significant but modest. 
The largest effect (a VDA effect size of $0.5687$) is obtained when using NMF with complete linkage.

\begin{table}[t]
\centering
\caption{Overall performance of the 16 configurations. LogCluster's baseline configuration is indicated in bold.}
\label{tbl:condensedsummary}\small
\vspace*{-1.5ex}
}
\rmfamily
\vspace*{-1.2ex}
\caption{Heatmaps of parameter sensitivity in terms of quality metrics for clustering pipelines using PCA, NMF and without dimensionality reduction (all with Complete linkage) and their differences. LogCluster is indicated by the black box.}
\label{fig:heat}
\vspace*{-2ex}
\end{figure*}

\begin{table}[b]\centering
\vspace*{-3ex}
\caption{Effect sizes (VDA) with respect to the baseline.}
\label{tbl:bestvbaseline}\small
\vspace*{-2ex}
%
\csvautobooktabular{best_against_baseline.csv}
\end{table}

\head{Parameter sensitivity} 
To analyze the sensitivity of the configurations against variations in the choice of parameters $\theta$ and $\gamma$, 
we draw heatmaps of the quality metrics AMI, Completeness and Homogeneity in Figure~\ref{fig:heat}.
We show data for the pipelines using PCA, NMF and without dimensionality reduction (all with Complete linkage), 
and the differences of respectively PCA and NMF with respect to the baseline (no dimensionality reduction). 
The performance of the original LogCluster approach is indicated by the black box. 
Higher values are indicated by darker tones. 
For the difference heatmaps, the positive values (in red) indicated where PCA resp.\ NMF outperform the baseline, 
the negative values (in blue) show where the baseline outperforms PCA resp.\ NMF.

\newcommand{\hmref}[1]{\ref{fig:heat}#1}

All heatmaps, except those for PCA (\hmref{b}, \hmref{g} and \hmref{l}), 
show that variations in the merge threshold $\theta$ have more impact 
than variations in the relative contrast $\gamma$.
When not using dimensionality reduction, 
$\theta$ must be in a particular range (roughly $[0.25 - 0.6]$) to achieve an adequate AMI score. 
The plots also shows that without dimensionality reduction, \emph{increasing} $\gamma$ is almost always beneficial: 
The score will either increase or stay the same. 
A few exceptions can be found for the very lowest values of $\theta$ (see especially the AMI scores in \hmref{c}). 
This effect is less pronounced for PCA and NMF.

Moreover, the heatmaps show how Homogeneity and Completeness are competing concerns. 
This is especially pronounced in the heatmaps without dimensionality reduction 
(\hmref{h} and \hmref{m}), where a high $\theta$ increases Completeness and decreases Homogeneity, and vice versa. 
We also see the impact of dimensionality reduction on the trade-off between Homogeneity and Completeness: 
No dimensionality reduction has a relatively stable performance for Completeness, while Homogeneity varies drastically,
especially as a function of $\theta$. 
PCA performs smoothly on both measures, but scores significantly better on Homogeneity than Completeness. 
NMF, on the other hand, achieves a consistent and high score on both measures.

Overall, both PCA and NMF show a more consistent performance across variations in $\gamma$ and $\theta$. 
PCA scores worse than no dimensionality reduction on Completeness (\hmref{f}), but very closely on Homogeneity.
Overall, no dimensionality reduction outperforms PCA in terms of mean AMI  
(shown by \hmref{a}, and confirmed in Table~\ref{tbl:condensedsummary}). 
Nevertheless, PCA scores a higher \emph{median} AMI than no dimensionality reduction when using Complete Linkage, 
and the Wilcoxon Signed-rank test established the two variants as significantly different.
NMF clearly beats no dimensionality reduction on all measures, and PCA on most (corroborated by Table~\ref{tbl:condensedsummary}).

\begin{figure}[t]\centering
\includegraphics[width=.95\columnwidth]{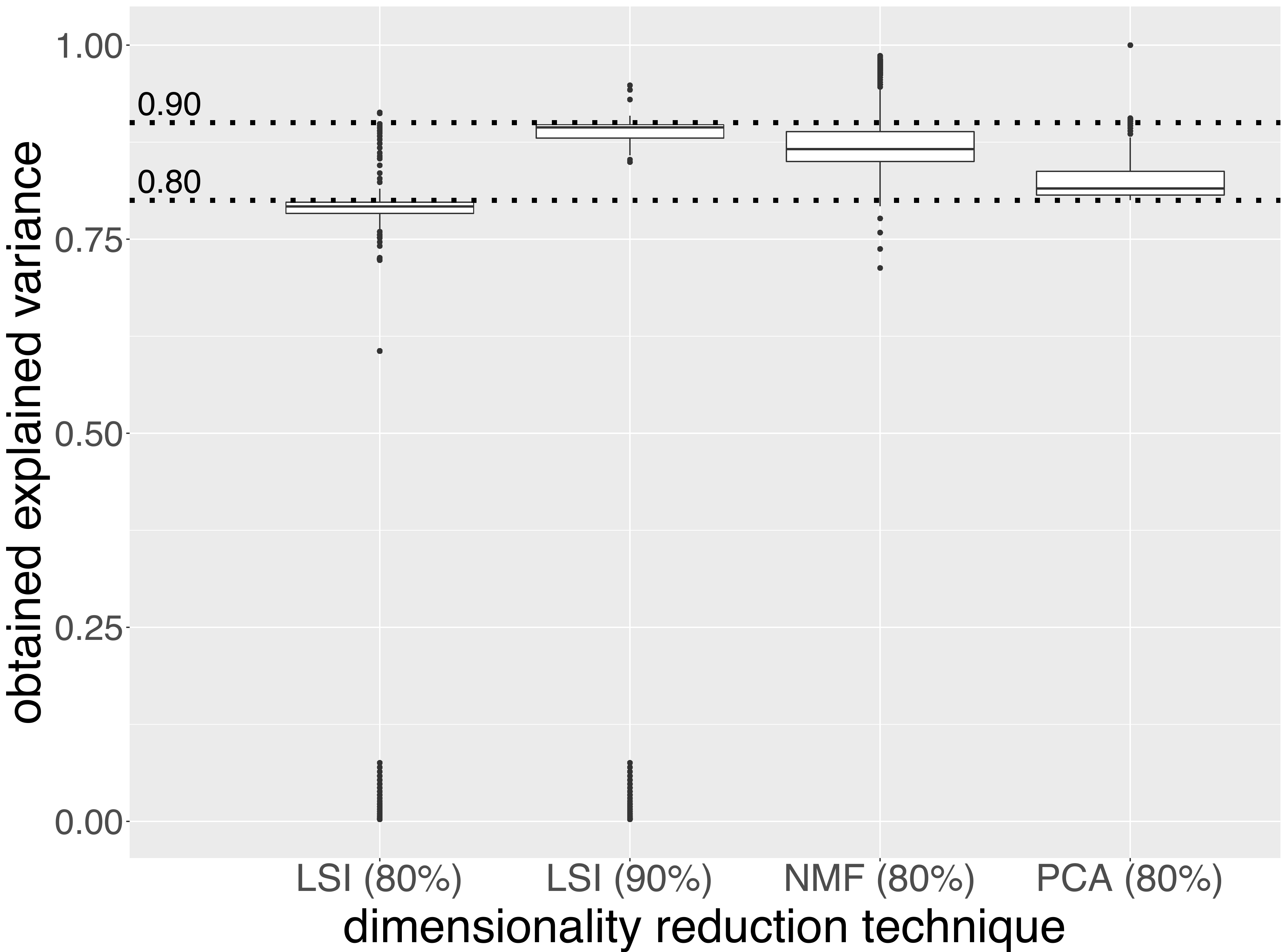}
\vspace*{-1ex}
\caption{Explained variance after dimensionality reduction.
The percentages in labels show desired explained variance.}
\label{fig:lsievar}
\vspace*{-2.5ex}
\end{figure}

\head{Analysis of LSI performance}
As shown in Table~\ref{tbl:condensedsummary}, LSI performs markedly worse than PCA and NMF. 
To investigate, 
we checked if the method for estimating the number of components (see Section~\ref{compestim}) could be to blame. 
We record what is the actual explained variance whenever dimensionality reduction is applied.
Figure~\ref{fig:lsievar} shows a boxplots of these. 
We see that a requested explained variance of 80\% always results in an actual explained variance of 80\% or higher for NMF and PCA. 
For LSI, however, the requested explained variance systematically is below the requested percentage.

We ran an additional experiment for LSI where we requested a 90\% explained variance. 
As seen in Figure~\ref{fig:lsievar}, this increased the actual explained variance, 
but again did not bring it above the requested threshold. 
This indicates that our PCA-based predictor of the number of components is not appropriate for LSI, 
which is somewhat unexpected, given that PCA and LSI are both based on Singular Value Decomposition. 
We leave the search for a better predictor to future work.

As was to be expected, increasing the requested explained variance to 90\% did increase LSI's median AMI scores to 
$0.31$ for Complete Linkage, 
$0.23$ for Weighted Linkage, 
$0.22$ for Average linkage, and
$0.029$ for Single Linkage. 
However, these still put LSI below the baseline in Table~\ref{tbl:condensedsummary} and (far) below PCA and NMF with corresponding merge criteria.
As such, our conclusions remain unchanged.

Another potential explanation for LSI's poor performance is that it is geared more towards natural language corpora, 
which easily have bag-of-words having 20 000 dimensions, 
and suffer from polysemy and synonymy~\cite{manning2008introduction}. 
Our largest bag-of-events has 850 dimensions (Table~\ref{tbl:dataset-summaries}), 
and it is unlikely that polysemy and synonymy play a role (unless there is an issue with log abstraction).
LSI may just fit better with problems whose dimensionality is at least an order of magnitude greater than what we observe.

\subsection{RQ2: impact of dimensionality reduction}

Overall, the impact of applying dimensionality reduction in the form of either PCA or NMF is increased robustness:
While the baseline scenario requires a sensible choice of $\theta$ for adequate performance, 
especially $NMF$ gives a much wider range of performant parameter choices, as seen in Figure~\ref{fig:heat}c and d.

However, the choice of dimensionality reduction strategy is crucial.
As discussed in the previous section, LSI performs worse than the baseline for all merge criteria.
While both PCA and NMF provide highly stable performance, the statistical analysis (Table~\ref{tbl:condensedsummary})
clearly establishes NMF as the superior choice (confirmed by Figure~\ref{fig:heat}). 

\head{Answer to RQ2} Our study suggests that using either NMF or PCA
yields more performant solutions that are more robust against changes in
$\gamma$ and $\theta$. NMF is clearly superior to PCA for the dataset
considered in this study, while LSI performs worse than the baseline.

\subsection{RQ3: impact of merge criteria}

The results presented in Table~\ref{tbl:condensedsummary} and Figure~\ref{fig:amiboxplots} show that 
Complete Linkage is the best performer for every choice of dimensionality reduction, 
while Single Linkage is unanimously the poorest performer. 
Weighted and Average Linkage show roughly similar performance.

The strong performance of Complete Linkage may indicate that our dataset has few outliers, 
as discussed in Section~\ref{complete}. 
The poor performance of Single Linkage may indicate that a \emph{chaining effect}, 
a known drawback of Single Linkage~\cite[Sec.\ 17.2]{manning2008introduction},
may be adversely affecting the results for that criterion. 
Our current study cannot conclusively establish whether this is indeed the case.

\head{Answer to RQ3} 
Our study suggests that using Complete Linkage yields the most performant solutions, 
which is consistent with the choice for this merge criterion in the work of Lin et al.~\cite{Lin2016} 
that was the starting point of our investigation.

\subsection{Threats to Validity}

\noindent
We identified the following threats that could affect the construct, internal, and external validity of our experimental results.

\head{Threats to Construct Validity}
The ground truth used for evaluation was established using
regular expressions handcrafted by our industrial collaborator to identify whether a log concerns a known issue.
One threat to validity is that these patterns do not cover all issues encountered in the full dataset
(which also motivates our research).
This is mitigated by creating the ground truth from the subset of data that is fully covered.
Another threat is that we can not guarantee or check the absence of matching errors in these patterns.
This is, to some extent, mitigated by the fact that they have been used to satisfaction by our partner.
A final threat to construct validity is that a hard clustering approach like LogCluster or its variants cannot
distinguish between multiple issues in a log file.
At best, a new cluster is formed of all logs sharing the same set of issues.
This is mitigated by limiting the dataset to logs that match with only one known issue.

\head{Threats to Internal Validity}
We implemented and thoroughly tested all algorithms and statistical procedures used in this paper in Python,
with the help of widely used libraries such as NumPy and SciPy.
However, we can not guarantee the absence of implementation errors which may have affected our evaluation.

\head{Threats to External Validity}
We evaluated the applicability and efficacy of the LogCluster algorithm and its variations on logs that
resulting from CDt activities at our industrial partner.
These logs varied considerably in size and events covered, which should provide a good picture of
the behavior that can be expected in various contexts.
However, we are likely not to have captured all possible variation,
and cannot rule out that different logging practices in other systems or organizations would lead to different results.

Moreover, as discussed in Section~\ref{sec:dataset},
we are left with a ground truth that is below half the size of the original dataset,
which limits the strength of statements that can be made about generalizability.

\section{Related Work}
\label{sec:related_work}

\noindent We distinguish the following categories of related work:

\head{Need for support to handle CE results}
The literature on adopting continuous software engineering practices,
such as CI, CDy, and CDt, frequently alludes to the challenges and needs
concerning systematic and integrated analysis of the wealth of data resulting from the automated build,
test, and deployment steps~\cite{Hilton2017,Laukkanen2017,Brandtner2015a,Brandtner2015b,Debbiche2014,Olsson2012}.

Hilton et al.~\cite{Hilton2017} interviewed developers from various industries on how they debug test failures detected in CI, and report: \emph{`` our participants told us that they get the output logs and start their search there. These output logs can be quite large in size though, [which] can create quite a challenge when trying to find a specific failure.''}

Shanin et al.~\cite{Shahin2017} identify the lack of transparency and awareness regarding test and build results 
as one of the main threats to adopting CDy,
together with the need for measures that improve coordination and collaboration on addressing such results.

These findings were echoed in discussions with our industrial collaborator \CiscoNorway{},
who developed an in-house tool to match CDt logs against hand-crafted regular expressions
to identify recurring issues.
The work described in this paper investigates to what extent this identification can be automatically supported, 
and what effort reductions can be achieved by clustering similar failing runs.
It removes the need for hand-crafting regular expressions by utilizing IR-based clustering techniques
and exploiting the contrast between events in passing and failing logs.

\head{Analysis of CE results}
Brandtner et al.\ propose SQA-Mashup~\cite{Brandtner2015a}, a quality awareness framework that integrates information from the entire CI-toolchain. They create a single service to monitor data that would otherwise be scattered over various locations.
In subsequent work, the same authors propose SQA-Profiles~\cite{Brandtner2015b}, a rule-based mechanism for the dynamic composition of CI dashboards based on stakeholder activities in the environment.

Nilsson et al.\ propose CiVIT~\cite{Nilsson2014a}, a technique to visualize testing activities to support CI.
They distinguish several types of testing and indicate the coverage for these types.
In addition, the scope and periodicity of various testing efforts are used to plot an integrated overall view of the testing status of a system.

St\aa{}hl and Bosch propose Cinders~\cite{Stahl2017b}, an architecture framework designed specifically to meet the needs of CI and CDy environments.
Based on an analysis of CI and CDy literature, they phrase twelve requirements that such an architectural framework should support. 
Cinders offers four separate viewpoints of the same underlying data model, 
with six optional layers of additional information which can be used to adjust the focus and level of detail within each of those viewpoints. 
It uses CiVIT~\cite{Nilsson2014a} to visualize the testing status.

These approaches all focus on the visual analysis of integrated CI results in dashboards to address the transparency and awareness needs discussed above.
The work that we present in this paper is complementary: 
We focus on integrating CI or CDy results into actionable clusters that can be treated more cost-effectively. 
Information derived from these clusters could in turn be incorporated in dashboards similar to the ones proposed by these authors.

\head{Analysis of crash reports} Several authors have worked on analyzing the crash reports that modern software systems offer to send to their developers upon detecting a runtime failure. 

Podgurski et al.~\cite{Podgurski2003} use a combination of supervised and unsupervised pattern classification to cluster software failure reports for prioritization and diagnosis. They first train a classifier to identify the features that distinguish crash reports from normal non-crashing behavior, and then use automated clustering on those features to classify the crash reports.
The initial classifier is used as a form of domain-specific dimensionality reduction.

Kim et al.~\cite{Kim2011} use supervised machine learning to learn the features of top crash reports from previous releases and predict the top potential crashes before a new release is made.

\head{Clustering of system logs}
The work presented in this paper explores to what extent techniques developed for the clustering of system logs can be used in a CE context.
Oliner et al.~\cite{Oliner2012} give an overview of the advances and challenges in system log analysis.

In Section~\ref{sec:background} we already presented an in-depth overview of the LogCluster approach by Lin et al~\cite{Lin2016} that forms the basis of our investigation. 
Shang et al.~\cite{Shang2013} propose an approach that uses the difference between small controlled runs and real-life data from a cloud environment to analyze Hadoop logs. This insight that the contrast between passing and failing runs can be exploited to improve clustering accuracy is also used in LogCluster. 

He et al.~\cite{He2016} present an experience report on using system log analysis for anomaly detection.
They describe six state-of-the-art log-based anomaly detection methods but do not draw conclusions on which method performs best.
Their focus on anomaly detection is different from ours, as we aim to group logs that fail for the same underlying issue in order to handle those issues more efficiently.

\head{Parameter Tuning}
Recent research highlighted that the successful application of machine learning and data mining algorithms 
in concrete domains is highly impacted by the selection of their configuration parameters~\cite{Maimon2010}.
We are not aware of any other work in a software engineering context that shares our goal of investigating the impact of 
particular dimensionality reduction techniques or clustering parameters such as the linkage criterion
on the quality of clustering software logs or crash reports.

\section{Concluding Remarks}
\label{sec:conclusion}

\subsection{Contributions} 

Our study considers the challenge of automatically grouping logs of runs that failed for the same underlying reasons,
so that they can be treated more effectively. 
We replicate and extend earlier work on clustering system log files~\cite{Lin2016} to assess its efficacy for the analysis of continuous deployment logs.
We consider optional inclusion of one of three dimensionality reduction techniques: 
Principal Component Analysis (PCA), Latent Semantic Indexing (LSI), and Non-negative Matrix Factorization (NMF). 
Moreover, we consider three alternative cluster merge criteria (Single Linkage, Average Linkage, and Weighted Linkage),
in addition to the Complete Linkage criterion used in the earlier work.
We empirically evaluate the 16 resulting configurations on continuous deployment logs provided by our industrial collaborator to answer the following three research questions:
\begin{compactenum}[(1)]
\item Can the application of LogCluster~\cite{Lin2016} be generalized to identifying problems in \emph{continuous deployment log files}?
\item To what extent does applying dimensionality reduction impact the results of  automated log clustering?
\item To what extent does the merge criterion in HAC impact  the results of automated log clustering?
\end{compactenum}
The study results allow us to answer these questions as follows:
\begin{compactenum}[(\text{\itshape ad}~1)]
\item The LogCluster approach generalizes to problem identification in continuous deployment logs.
\item Including an NMF-based dimension reduction step results in significantly better overall performance and robustness.
\item Complete Linkage performs best of all the merge criteria that were considered.
\end{compactenum}
\head{Conclusion} 
We conclude that problem identification via automated log clustering is improved by including dimensionality reduction,
as it decreases the pipeline's sensitivity to parameter choice, 
thereby increasing its robustness for handling different inputs, 
a desirable characteristic for an unsupervised approach such as automated log clustering~\cite{Aggarwal2015}.

\subsection{Future Work}
We see several directions in which this work can be extended. 
First and foremost, these techniques should be evaluated on a wider range of case studies. 
A challenge in that respect is establishing a good ground truth for the qualitative evaluation of results, 
as fully labelled sets of log files are rare. 
One option to address this challenge could be to programmatically synthesize labelled log files with known characteristics.

Another area for future work is in finding better predictors for the number of components used by LSI, and possibly NMF. 
Although our intuition to use the amount needed by PCA as predictor for NMF worked well, 
it underperformed for LSI, even though both PCA and LSI are based on singular value decomposition.                   

A final direction for future work is (support for) the in-depth analysis of the cluster merging behavior using
dendrograms and merge distances. 
Such an analysis could help understand why some of the merge criteria (like Single Linkage) performed poorly, 
and why others (such as Complete Linkage) performed well, despite their known pitfalls.

\begin{acks}
We thank Marius Liaaen and Thomas Nornes of Cisco Systems Norway for extensive discussions, 
help with obtaining and understanding the data set, and for developing the log abstraction
mechanisms.
This work is supported by the \grantsponsor{RCN}{Research Council of Norway}{https://www.rcn.no} through the
Certus SFI (\grantnum{RCN}{\#203461/030)}.
\end{acks}

\printbibliography 

\end{document}